# Method of targeted delivery of laser beam to isolated retinal rods by fiber optics

**Nigel Sim,[1] Dmitri Bessarab,[2] C. Michael Jones,[2] and Leonid Krivitsky[1,*]**

[1] *Data Storage Institute, Agency for Science Technology and Research (A-STAR), 117608 Singapore*
[2] *Institute of Medical Biology, Agency for Science Technology and Research (A-STAR), 138648 Singapore*

*[*]Leonid_Krivitskiy@dsi.a-star.edu.sg*

**Abstract**: A method of controllable light delivery to retinal rod cells using an optical fiber is described. Photo-induced current of the living rod cells was measured with the suction electrode technique. The approach was tested with measurements relating the spatial distribution of the light intensity to photo-induced current. In addition, the ion current responses of rod cells to polarized light at two different orientation geometries of the cells were studied.

______________________________________________________________

**References and links**

[1] J.E. Dowling, *The Retina* (Harvard University Press, Cambridge, Massachusetts, 1987).
[2] F. Rieke, and D. A. Baylor, "Single-photon detection by rod cells of the retina," Reviews of Modern Physics **70**, 1027-1036 (1998).
[3] S. Hecht, S. Shlaer, and M.H. Pirenne, "Energy, quanta and vision," J. Gen. Physiol. **25** (6), 819-840 (1942).
[4] D.A. Baylor, T.D. Lamb, and K.-W. Yau, "The membrane current of single rod outer segment," J. Physiol. **288**, 589-611 (1979).
[5] D.A. Baylor, T.D. Lamb, and K.-W. Yau, "Responses of retinal rods to single photons," J. Physiol. **288**, 613-634 (1979).
[6] P.B. Detwiler, J.D. Conner, and R.D. Bodoia, "Gigaseal patch clamp recording from outer segments of intact retinal rods," Nature **300**, 59-61 (1982).
[7] A.L. Hodgkin, P.A. McNaughton, B.J. Nunn, and K.W. Yau, "Effect of ions on retinal rods from *Bufo marinus*," J. Physiol. **350**, 649-680 (1984).
[8] E.N. Pugh, Jr., and T.D. Lamb, "Phototransduction in vertebrate rods and cones: molecular mechanisms of amplification, recovery and light adaptation," Chapter 5, 183-255, in Handbook of biological physics, **3**, *Molecular Mechanisms of Visual Transduction*, edited by D.G. Stavenga, W.J. de Grip, and E.N. Pugh, Jr. (Elsevier science, 2000).
[9] W.H. Xiong, and K.-W. Yau, "Rod sensitivity during *Xenopus* development," J. Gen. Physiol. **120**, 817-827 (2002).
[10] J. L. Schnapf, "Dependence of the single photon response on longitudinal position of absorption in toad rod outer segments," J. Physiol. **343**, 147-159 (1983).
[11] V. I. Govardovskii, D. A. Korenyak, S. A. Shukolyukov, and L. V. Zueva, "Lateral diffusion of rhodopsin in photoreceptor membrane: a reappraisal," Molecular Vision **15**, 1717-1729 (2009).
[12] D. A. Baylor, and R. Fettiplace, "Light path and photon capture in turtle photoreceptors," J. Physiol. **248**, 433-464 (1975).
[13] F. I. Harosi, "Absorption spectra and linear dichroism of some amphibian photoreceptors," J. Gen. Physiol. **66**, 357-382 (1975).
[14] D.A. Baylor, B.J. Nunn, and J.L. Schnapf, "The photocurrent, noise and spectral sensitivity of rods of the monkey *Macaca fascicularis*," J. Physiol. **357**, 575-607 (1984).
[15] G. Horvath, and D. Varju, *Polarized Light in Animal Vision*, (Springer-Verlag, Berlin, Heidelberg, 2004).
[16] N. W. Roberts, and M. G. Needham, "A mechanism of polarized light sensitivity in cone photoreceptors of the goldfish *Carassius auratus*," Biophys. J. **93**, 3241-3248 (2007).
[17] F. L. Tobey, Jr., and J. M. Enoch, "Directionality and waveguide properties of optically isolated rat rods," Investigative Ophthalmology **12**, 873-880 (1973).
[18] A. Koskelainen, K. Donner, G. Kalamkarov, and S.Hemila, "Changes in the light-sensitive current of salamander rods upon manipulation of putative pH-regulating mechanisms in the inner and outer segment," Vision Res. **34**, 983-994 (1994).
[19] W.A. Shurkliff, *Polarized Light,* (Harvard University Press, Cambridge, Massachusetts, 1962).

[20] T.D. Lamb, P.A. McNaughton, and K.-W. Yau, "Spatial spread of activation and background desensitization in toad rod outer segments," J. Gen. Physiol. **319**, 463-496 (1981).
[21] A. Dhatak, K. Thyagarajan, *Introduction to Fiber Optics,* (Cambridge University Press, Cambridge, UK, 1998).

## 1. Introduction

The vertebrate retinal rod cells represent advanced low-light detectors granted by nature. They have the ultimate sensitivity, low dark noise, and an extremely small foot-print. For example, the amphibian rod cell of about 5*50 micron size represents a self-contained photo-detector, which has a light-sensitive element (rhodopsin pigment) along with a "built-in" chemical power supply (ATP produced by mitochondria). The isomerization of the rhodopsin molecules embedded in membranous disks of the rod outer segment (ROS) by the impinging light triggers a chain of reactions, referred to as photo-transduction, which causes the closure of definite ionic channels in the cell membrane [1, 2]. The closure leads to a net reduction of an inward ion current and can subsequently be translated into a measurable electrical readout.

One of the early studies of the response of visual systems to faint light stimuli were conducted by Hecht *et al.* back in the 1940's using the "frequency of seeing" experiments, where light flashes were presented to the dark-adapted human subject [3]. Subsequently, photocurrent recordings from isolated retinal rods of various species became accessible with the introduction of the suction electrode technique and eventually with the patch clamp method [4-6]. The success of both suction electrode and patch clamp techniques allows studying many different aspects of light perception, including development, dark adaptation, and others [7-9].

As rod cells *in vivo* are optimized to operate at low light intensities, an accurate characterization of cell response to a given number of impinging photons requires a precise control and optimization of light delivery to the cell. In the majority of earlier studies, the light source (typically a broadband lamp with a set of band-pass filters) impinging on a retinal rod cell was positioned above the cell such that the light beam passing through projected vertically and nearly perpendicular to the long axis of the rod cell [4,5,10,11]. However, such a configuration, shown in Fig. 1a, and further referred to as *transverse configuration*, does not provide efficient light capture for several reasons [12].

First, since the impinging light has a short travelling path in the ROS (it crosses the cell along its diameter), the effective absorption length is considerably smaller in comparison with the case when the light travels along the cell's long axis. Second, since the dipole moments of rhodopsin molecules are constrained within the plane of the disc membrane, the light capture by the cell is highly sensitive to the polarization of the impinging light [13-16]. And third, individual rods in the retina physiologically absorb axially incident light [16].

In this work we suggest a fiber optics based method for the precise delivery of light to the retinal rod cells. The use of fiber optics allows the delivery of light stimuli to the rod with full control over the stimuli intensity profile, spatial location, and polarization. In the experiments reported below, we tested the rod response by monitoring a photo-induced current when the light stimulus was delivered parallel to the long axis of the rod cell, a configuration referred to as *axial configuration,* and shown in Fig. 1b. This method simply mimics the pathway of light impinging on retinal rods with their long axis pointing towards the pupil of the eye and optimizes photon capture in comparison with the conventional approach described above. Thus, in the axial configuration, the light traveling path in the cell is 7-9 times longer than in the case of the transverse configuration. Moreover, the light capture is assisted by the guiding properties of the rod resulting from the differences in refractive indices between the rod and the surrounding solution [12, 17]. Furthermore, as the polarization plane of light emerging from the tip of the fiber, is always orthogonal to the direction of its propagation, it will always be in plane with the membrane discs of the ROS. Thus, the cell response will be much less sensitive to stimuli polarization [13].

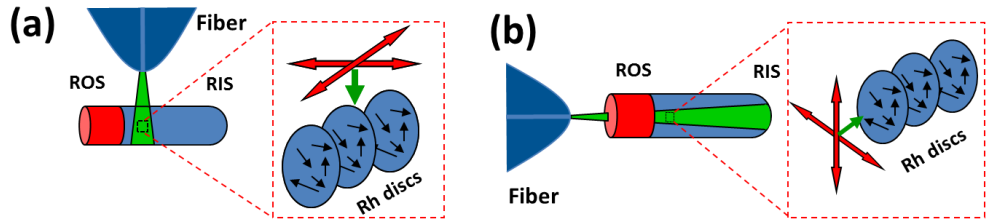

Fig. 1. Transverse (a) and axial (b) configurations for rod illumination. ROS and RIS, stand for outer and inner segment of the rod, respectively. The green stripe at the ROS shows a light absorption region. The insets show two orientations of light polarizations (red arrows) with respect to the membrane discs of the ROS, and the green arrow is a wave vector of light. The black arrows are orientations of dipole moments of rhodopsin molecules within the membrane discs [15].

## 2. Methods

*2.1. Cell preparation*

All manipulations with animals were made according to the regulations of IACUC at the Institute of Medical Biology (IMB). The adult male frogs (*Xenopus laevis*) were dark-adapted overnight and sacrificed in accordance with Standard Operating Procedure for IMB Aquatic Facility. The eyes were enucleated and hemisected under a dissecting microscope with infra-red LED illumination (850 nm, bandwidth 20 nm) within one to two hours after the sacrifice. The retina was gently detached and separated from the underlying retinal pigment epithelium and cut into pieces in a Ringer solution to release individual rod cells. The Ringer containing 111 mM NaCl, 2.5 mM KCl, 1.6 mM $MgCl_2$, 1.0 mM $CaCl_2$, 10 mM D-glucose, 3 mM HEPES was adjusted to pH 7.8 with NaOH [4] and enriched with 5% v/v L-15 Leibovitz medium (all chemicals were purchased from Sigma-Aldrich) [18]. The experiments were conducted at room temperature 22-24$^0$C. The selected rods remained functional for consistent recordings for up to 90 - 110 minutes.

*2.2. Electrophysiological recordings*

The rod cells were loaded in a shallow Ringer-filled 25 mm diameter chamber (Siskiyou, Inc.) of an inverted microscope (Leica, DMI 3000B) placed in a light tight enclosure. The preparation was observed with a monochrome CCD camera (Leica, DFC340FX) under low intensity infra-red LED illumination (920 nm, bandwidth 25 nm). The cells were perfused with fresh Ringer solution at the flow rate of 0.5 ml/min. The solution was grounded by an Ag/AgCl reference electrode through a 3.5% 100 mM KCl agar bridge. A patch clamp amplifier (HEKA, EPC10) was used to record the light-induced photocurrent. Intact single rods were selected and were positioned in a bended tight fitting glass micropipette with their ROS in the pipette. The initial resistance of the glass micropipette was about 520-550 KOhm and it grew by 10-12 times once the rod was fitted into the pipette tip. The functionality of the cell was confirmed by regular observations of the amplitude of the cell response (typically of 20-25 pA) to brief light flashes of saturating intensity. The tip of the pipette was bent to an angle of 45±5° (135° in Fig.2b) at a distance of 4-5 mm using an open flame burner. Initially, the pipette tip was pointing downwards at the surface of the cover slip of the chamber, where the rods had settled over time. After the rod was constrained, the pipette was manually tilted in the direction indicated by the dashed arrow in Fig. 2b, so that the bended tip was oriented parallel to the cover slip. The pipette, with a constraint cell, as well as the optical fiber tip (see below) were mounted on two micromanipulators (Sutter Instruments, MPC-200), which provided an accurate XYZ-spatial positioning with a resolution of 0.3 μm (Fig. 2c).

*2.3 Optical setup*

The beam of a continuous wave (cw) frequency-doubled Nd:YAG laser (Photop) at 532 nm wavelength, and an optical power 30mW, was chopped by an optical shutter (Melles Griot) producing pulses with the duration of 30 ms and a repetition rate of 15 s (Fig. 2a). The laser pulses were strongly attenuated by a variable neutral density filter (NDF, Thorlabs) and divided by a 50/50 non-polarizing beam-splitter (BS, Thorlabs). The pulse in the transmitted arm of the BS was used to stimulate the cell, whilst the equivalent reflected pulse was used to control the intensity of the stimulus. In both arms, the light was coupled into single mode optical fibers (SM) of 2 m length (Nanonics, model: 460-HP, cut-off wavelength at 430±20 nm, numerical aperture 0.12) by two aspheric lenses (L, Thorlabs) with $f$=6.24 mm. The fiber, which was used to stimulate the cell, had a tapered tip, which was channelled through a plastic pipette bent at 45°. In the reflected arm of the BS the FC-terminated fiber tip was connected to a single photon avalanche photodiode (Perkin Elmer, SPCM-AQRH-15-FC, quantum efficiency of 50% at 532 nm, dead time 32 ns) which reads the photon flux in real time. A careful adjustment of fiber launcher stages and coupling lenses provided that the number of photons emitted from both of the fibers and measured by the APD was equal.

In the experiment, two orientations of the optical fiber tip relative to the rod cell in the recording pipette were tested. In the *transverse configuration*, the optical fiber tip was positioned vertically above the preparation, similar to the approach described earlier [4, 10, 14]. In the *axial configuration*, the optical fiber tip was carefully manoeuvred using the micromanipulator to be aligned along the long axis of the rod cell. An optimal alignment was confirmed by observation of the maximum response to repetitive light pulses at fixed intensity.

To study the polarization dependence of the cell response, a zero-order half-, quarter- and another half-wave plate (HWP1, QWP, HWP2, Dayoptics) were introduced in a sequence to compensate for polarization rotation in the optical fiber (Fig. 2a). The initial alignment of the setup was assisted by a polarization analyser, placed perpendicular to the beam exiting the fiber tip (not shown). The NDF was removed during the alignment and the light intensity was controlled with an optical power meter (Thorlabs, Model PM100D with S120C sensor). Both, QWP and HWP2 were rotated to maximize light intensity passing through the analyser, whilst the orientation of HWP1 was fixed at zero degrees. After the pre-alignment, the analyser was removed and the orientations of QWP and HWP2 remained fixed. The transformation from the initial polarization to the orthogonal one was achieved through a 45° rotation of HWP1 in 10-15° increments [19]. It was provided, that in the absence of mechanical and thermal disturbances of the fiber the polarization at the output remained stable for at least 5-6 hours.

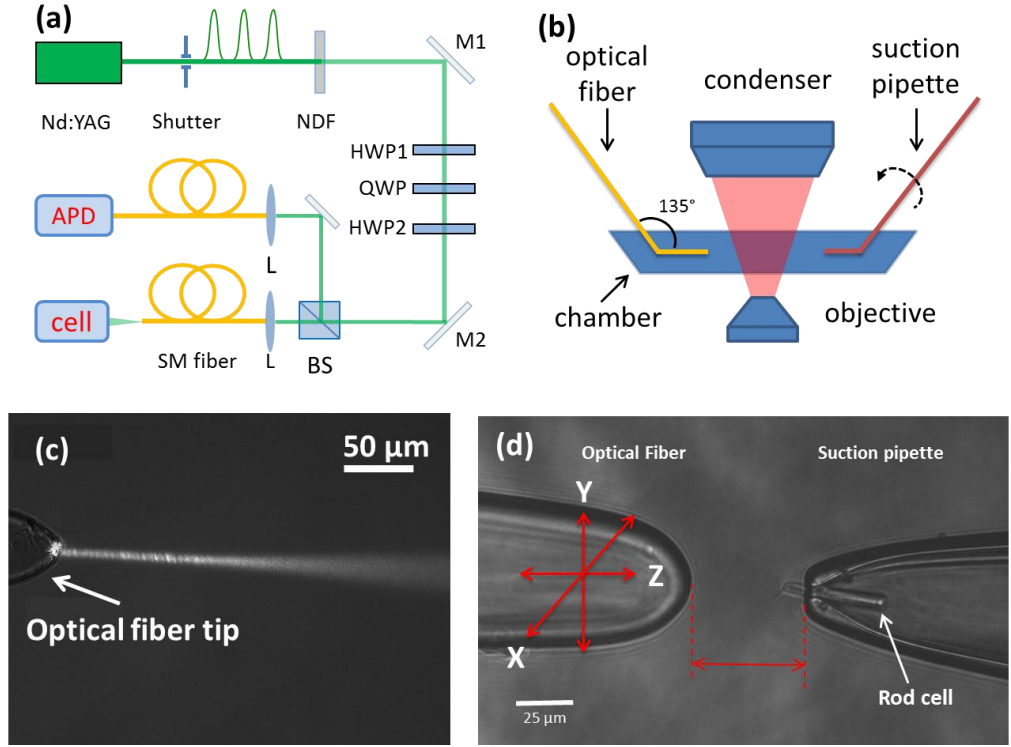

Fig. 2. (a) Optical layout. A cw 532nm Nd:YAG laser is chopped by a shutter and attenuated by a variable neutral density filter (NDF). Half-, quarter- and another half-wave plate (HWP1, QWP, HWP2) are used to control the polarization. A 50/50 non-polarizing beam-splitter (BS) splits the beam into two beams, and lenses (L) focus them into two single mode (SM) fibers. The intensity of the stimulus is measured with a single photon avalanche photodiode (APD). (b) Arrangement of the bent suction pipette and the bent tip of the optical fiber at the microscope chamber in the axial configuration (side view). Dashed arrow shows the tilt direction of the suction pipette from the initial position with the tip pointing downwards to the axial configuration. (c) Side-view of the 532 nm laser beam profile collimated by the rounded fiber tip (fiber lens) in a Ringer solution (top view). (d) Microscope image of the tapered fiber tip and the suction pipette with a constrained retinal rod (top view). The fiber tip is scanned in XYZ-directions, shown by red arrows. The horizontal arrow between two vertical dashed lines shows the distance between the fiber tip and the pipette.

## 3. Results and discussion

*3.1. Intensity dependence of the cell response*

First, by varying the transmission of the NDF, the dependence of the cell response on the number of photons emitted from the fiber was measured in the axial configuration at the distance of 40 μm between the fiber tip and the suction pipette. The waveforms at different attenuations of the photon flux are shown in Fig. 3a and their normalized amplitudes, plotted versus the normalized number of impinging photons per pulse measured by the APD, are shown in Fig. 3b. The dependence of the response amplitude $A$ on the number of photons $n$ in Fig. 3b was fitted by Michaelis equation (solid line) $A(n)=A_0(1-exp(-Bn))$ using a Levenberg-Marquardt algorithm (Origin Lab) [20]. The fitting with the following parameters $A_0$=21pA and $B^{-1}$=3650 photons/pulse resulted in a coefficient of determination (COD)=0.998. The dependence was normalized by the maximum amplitude $A_0$ and by the number of photons corresponding to a half saturating amplitude $n_0$=2500 photons/pulse. Subsequent measurements of the intensity profile (Section 3.2) and the polarization dependence of the response (Section 3.3) were conducted within the linear region of the cell response, i.e. at the photon fluxes producing amplitudes which do not exceed 1/3 of the saturating amplitude $A_0$.

Note that the number of photons in the stimulus beam plotted in the abscissa of Fig. 3b is a result of a direct measurement by the APD. Thus, the *light coupling efficiency*, introduced as a ratio between the amplitude of the cell response (measured within the linear region of the cell response curve) and the corresponding number of photons per pulse, emitted from the fiber, can be directly estimated. The typical value of the light coupling efficiency for the axial configuration was about 10-13 fA/photon. In contrast, for the transverse configuration under similar experimental conditions, the efficiency was not exceeding 0.05-0.1 fA/photon. Such a difference in values of coupling efficiencies is due to an optimized spatial overlap between the light spot and the cell, and a longer cell absorption length in the case of the axial configuration.

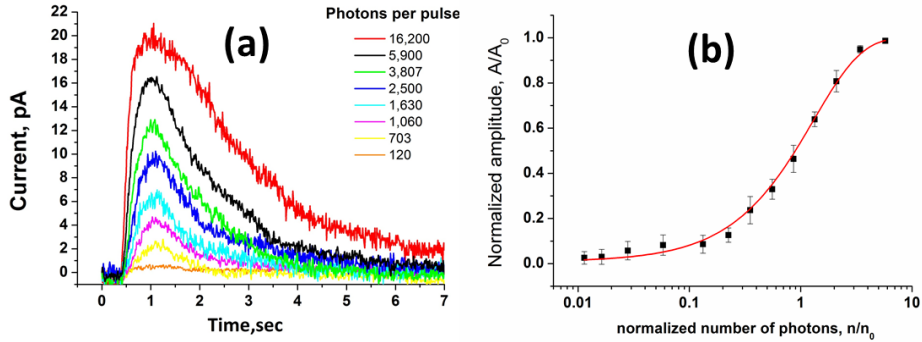

Fig. 3. (a) Waveforms of the cell response at different number of impinging photons in the axial configuration. The distance between the fiber tip and suction pipette is 40 μm. The color legend indicates the corresponding number of impinging photons. (b) Dependence of the normalized amplitude of the cell response in (a) on the number of impinging photons, measured by the APD. The amplitude is normalized by $A_0$=21 pA, and the number of photons is normalized by $n_0$=2500 photons/pulse which corresponds to a half saturating amplitude of 10.5 pA. The solid line is a fit by Michaelis function. Each waveform and plotted point in (a,b) is an average of 20 responses.

### 3.2. Spatial scan of light intensity profile

The diffraction of the Gaussian beam emitted into free space from the fiber tip causes its divergence in the direction of the propagation [21]. The experimental scan of light intensity emitted from the fiber was performed in the axial configuration. For this purpose, the tip of the fiber was scanned in the spatial plane orthogonal to the long axis of the cell (X-, Y- directions in Fig. 2d). The obtained profiles of the amplitude of the cell response in Y- direction at different distances between the fiber tip and the pipette are shown in Fig. 4. For easier comparison of the results, the amplitudes in each experiment were normalized by the maximum achievable value. The zero displacement corresponds to the axial alignment of the fiber tip with the cell. The results clearly demonstrate a Gaussian shape of the intensity distribution, which corresponds to a single Gaussian spatial mode which propagates in the fiber. The full width on the half maximum (FWHM) grows with the increasing distance between the cell and the fiber tip with the corresponding values listed in Fig. 4. The values of FWHM were found to be in agreement (within 7-10%) with the independent experimental test of the light intensity profile in the Ringer solution, observed by the CCD in a linear regime (insets in Fig. 4), taking into account the transverse diameter of the rod cell of 4 μm. Solid lines in Fig. 4 are the Gaussian fits of the experimental data with the same algorithm as the one used in Section 3.1, which resulted in COD=0.97-0.99. Similar dependences were obtained for the scans in the perpendicular direction (X-direction).

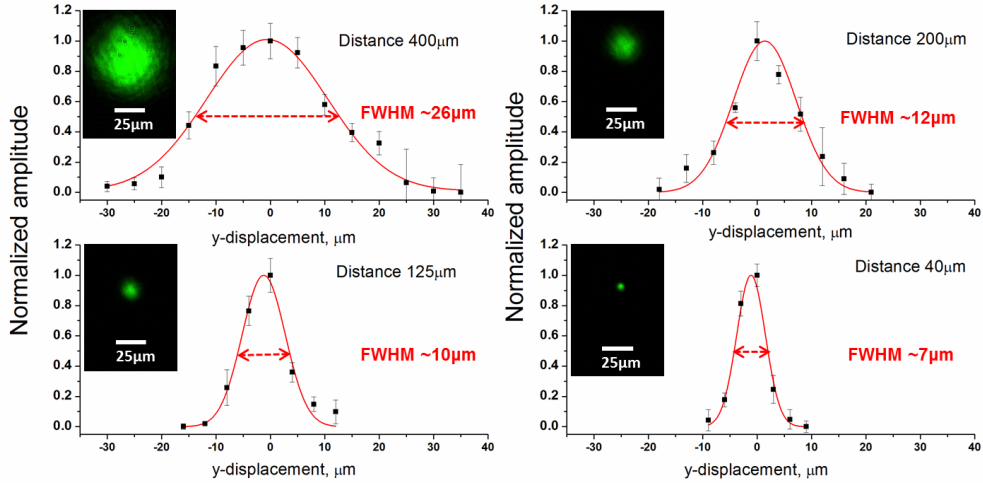

Fig. 4. Results of a transverse scan of light intensity in Y-direction at different distances between the pipette and the fiber tip, represented by Z-direction in Fig. 2d. Zero displacement corresponds to the axial alignment of the cell and the fiber tip. Plotted points are amplitudes of the cell response normalized by the maximum achievable value (average of 15 responses with 3 different cells) and red solid lines are experimental fits with Gaussian curves. Insets show transversal beam profiles measured by the CCD at the corresponding distances from the fiber tip.

### 3.3. Polarization dependence of the cell response

The cell response to linear polarized light was tested in both the transverse (Fig. 5a) and the axial (Fig. 5b) configurations. In the experiment, the initial polarization (HWP1 at 0°) was set to be perpendicular to the long axis of the cell. It was transformed to the orthogonal linear polarization (along the long axis of the cell) by a 45° rotation of HWP1.

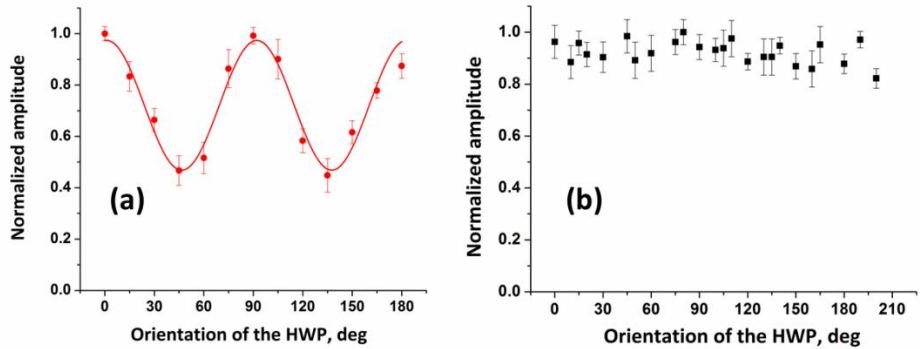

Fig. 5. Polarization dependence of the normalized cell response in transverse (a) and axial (b) configurations to linear polarized light. Solid line in (a) is the fit of the experimental data with COD=0.98. Each experiment was repeated for 5 different cells. At each given orientation of the HWP1 the response was averaged over 20 light pulses. Average amplitude of the cell response along with its standard deviation was normalized by the maximum response amplitude achievable for a given dependence.

In the transverse configuration, the obtained results clearly show a sinusoidal dependence of the response amplitude to changes in the polarization of the impinging light (Fig. 5a). The results are fitted by the following function: $Y(\theta)=1-Y_0 sin^2(2\theta+\varphi)$, where $Y_0$ is the fringe amplitude, $\theta$ is the orientation of the HWP1, and $\varphi$ is a phase shift. In the experiment with a linear polarized light, (see Fig. 5a) the *visibility (V)* of the dependence, defined as $V=(Y_{max}-Y_{min})/(Y_{max}+Y_{min})$, where $Y_{max}$ ($Y_{min}$) is the maximum (minimum) value, was $V=38.6\pm3\%$, and the phase shift $\varphi=-0.1\pm0.04°$. The corresponding dichroic ratio (DR), introduced as $DR=Y_{max}/Y_{min}$, is equal to $2.26\pm0.2$. The obtained result agrees with the range of values reported in [14] with the suction electrode technique, however it is slightly smaller than the corresponding value obtained in microspectrophotometry studies DR=3.12-3.4 [13, 16]. The difference may be attributed to slight misalignment of the polarization basis and the orientation of the cell axis.

The modulation of the response is much less pronounced in the case of the axial configuration (Fig. 5b). Illumination with linear polarized light resulted in $V=3.8\pm1.2\%$ and the corresponding $DR=1.08\pm0.04$. We checked that depolarization of light due to propagation through the Ringer solution was negligible. The obtained results are in a good agreement with Roberts & Needham's results from the absorbance measurements, which implied that axially incident polarized light will be absorbed isotropically in rods of the species which are not known to exhibit polarized light sensitivity [16].

## 4. Conclusions

We have introduced a fiber assisted light interface with retinal rod cells and tested it by measuring the spatial distribution of light intensity and polarization dependence of the cell response in transverse and axial coupling configurations. For the particular case of the axial configuration, inspired by the preferred route that light impinges on the photoreceptor in nature, light capture can be greatly enhanced due to the longer travelling path within the cell, the better match between the light spatial profile and the cell shape, and its invariance to polarization.

In our approach the efficiency of light delivery, introduced as a ratio between the amplitude of the cell response and the number of photons per pulse emitted from the fiber, can be directly estimated. We suggest that the cell response to known number of stimulating photons at given conditions could serve as a quantitative characteristic specific to the visual processes in the photoreceptor tested. Taken together with the axial configuration of illumination of the photoreceptor by the optical fiber, this potentially allows conducting comparative studies of photoreceptors between different vertebrate species.

The demonstrated capability of three dimensional characterization of the intensity profile in the vicinity of the fiber tip allows optimization of light delivery efficiency into retinal rods. These techniques and results should prove relevant to development of future applications of optical fibers in stimulating retinal cells such as (but not limited to) studies of the directional and spatial dependence of light capture, light guiding properties of retinal rods, polarization effects.

## Acknowledgements

We would like to acknowledge Yuen Wai Hong, Tan Jun Xian, Wong Sze Kai and Vadim Volkov for their help with the experiments and the discussions. This work was supported by A-STAR Joint Office Council grant (No. 10/3/EG/05/03) and A-STAR Investigatorship grant.